\documentclass[aps,floats,amssymb,amsmath,prb,onecolumn]{revtex4}
\usepackage{calc}
\usepackage{psfrag}
\usepackage{graphicx}

\begin{document}

\title{Towards the Hall viscosity of the Fermi-liquid-like phase at the filling factor of 1/2}

\author{M. V. Milovanovi\'c}
\affiliation{ Scientific Computing Laboratory, Institute of Physics,
University of Belgrade, P. O. Box 68, 11 000 Belgrade, Serbia}

\begin{abstract}
We discuss the Berry curvature calculations of the Hall viscosity
for the unprojected to the lowest Landau level wave function of the
Fermi-liquid-like state. We conclude, within assumptions made, that
in the linear response, with small deformation of the system and in
the thermodynamic limit, the Hall viscosity takes the value
characteristic for the Laughlin states. We present arguments that
the value is the same even for general deformations in the same
limit.
\end{abstract}


\date{\today}

\maketitle

\section{Introduction}

The Hall viscosity may represent an additional invariant by which we
can characterize quantum Hall states. In the work of Read
\cite{ReadV} it follows directly from conformal field theory (CFT)
description of quantum Hall states and it is related to the
conformal spin of the CFT field that is associated with the electron
description. In this way some non-unitary CFTs and corresponding,
presumably gapless, quantum Hall states may have a well-defined Hall
viscosity. The question is whether the ``quantization" of the Hall
viscosity will persist as we modify model Hamiltonians for which the
states are exact zero energy states i.e. whether these phases
despite being gapless have an invariant - the Hall viscosity. We
still do not have a definite answer for gapfull cases, but this
question seems related and relevant. The second question is: if a
quantum Hall state cannot be expressed by a CFT correlator like the
Fermi-liquid-like state at 1/2, whether it may still have the
invariant - Hall viscosity equal to the value of Laughlin states
which simple considerations of orbital spin of electron may motivate
\cite{ReadV}. On the other hand in that case we may think that the
Hall viscosity is the limiting value of the series of Hall
viscosities of the states of Jain series that leads to the
Fermi-liquid-like state. If the Hall viscosity is proportional
\cite{ReadV} to the shift of the state multiplied by its filling
factor the limiting value will diverge and certainly we will not
have a finite (``quantized") value associated with other states
constructed as CFT correlators.

In this paper we will discuss the Hall viscosity of the
Fermi-liquid-like state \cite{rzr} and phase \cite{hlr} at $\nu =
1/2$. We will consider the unprojected to the lowest Landau level
(LLL) wave function for the ground state introduced in Ref.
\onlinecite{rzr}, but at $\nu = 1$ i.e. the Fermi-liquid-like state
of bosons \cite{readbos}. The bosonic and  fermionic state share the
same underlying Fermi-liquid-like physics. The bosonic version will
facilitate the calculations of the Berry curvature and its
coefficient, the Hall viscosity, in  the case of these gapless
systems. In the introductory Section II we will review the basic
ansatz for calculating the Hall viscosity in the case of quantum
Hall states that was used in Ref. \onlinecite{TVjcm}. In Section III
the Hall viscosity of free Fermi gas is discussed as a step towards
the calculation for the Fermi-liquid-like state. In Section IV the
Hall viscosity as a response to a small deformation of the
Fermi-liquid-like state is discussed (a) when the system is
quasi-one-dimensional, (b) in the case when the Fermi surface is
rectangular, and finally (c) in the case of interest i.e. when the
Fermi surface is isotropic and circular. The next section, Section
V, discusses the Hall viscosity of a system under a general
deformation, and Section VI the importance of the inversion symmetry
for the neutral part as an effective symmetry for composite fermions
(CFs) that is present in the systems with rectangular shape, which
are deformed. Section VII contains a discussion of results and
conclusions.

\section{Hall viscosity of quantum Hall states}

The approach \cite{Avron,ReadV,TV,lev} to the Hall viscosity that we
find in the literature relies on calculating the Berry curvature of
shear deformations of the ground state that is adiabatically
transformed. This means an assumption is made that the state is
non-degenerate along the process. This can be assured if the system
has a gap which is the characteristic of usual quantum Hall states.
The shear deformations are examined by following how the quantum
liquid is spread out in the deformed geometry of a torus (we will
discuss the boundary conditions (BCs) shortly)- see Figure 1.
\begin{figure}[htb]
\centering
\includegraphics[scale=0.4, angle=0]{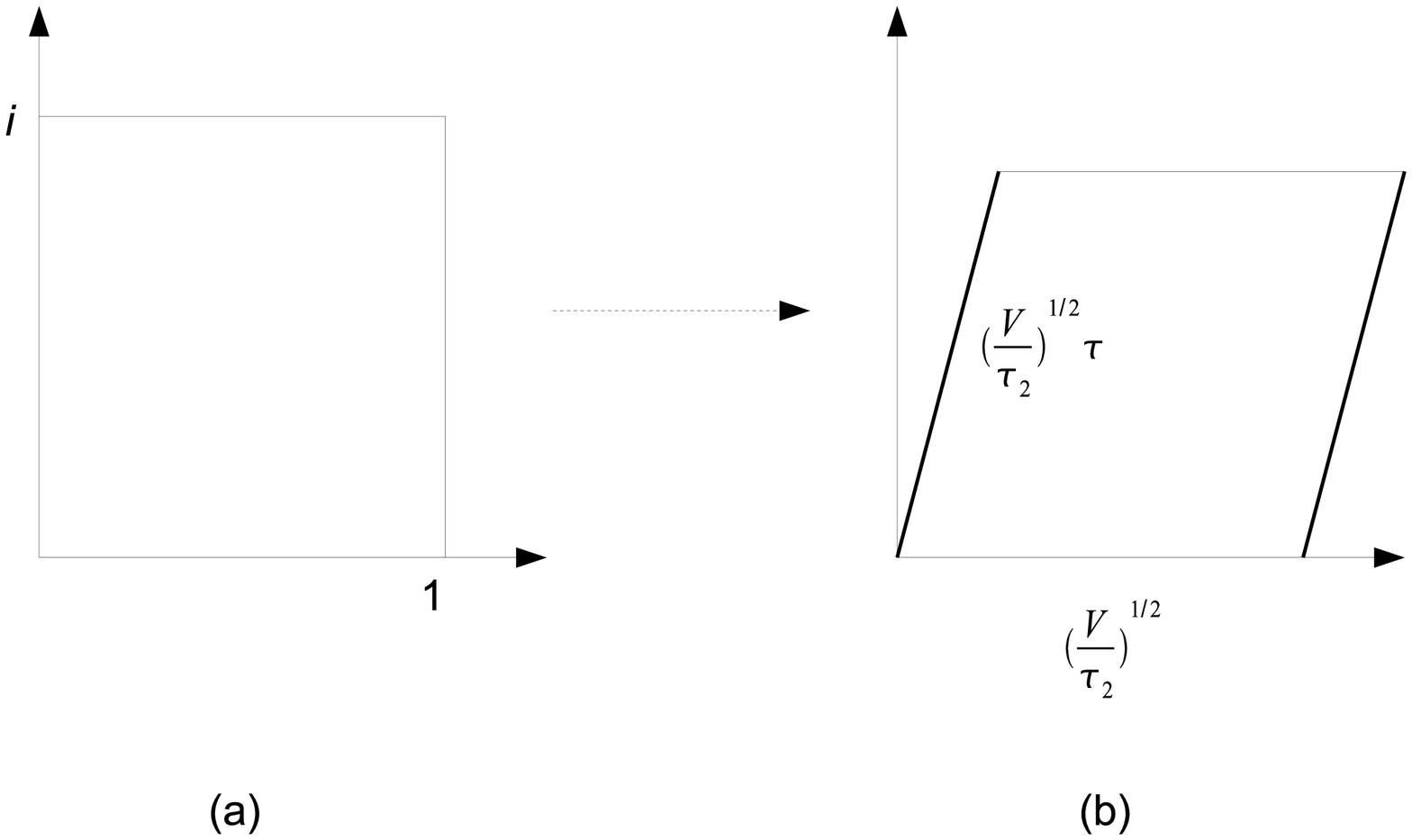}
\caption{Deformed torus} \label{fig1}
\end{figure}

The parameters $V$ and $\tau = \tau_1 + i \tau_2$ describe the
deformation from the reference point $V = 1$ and $\tau = i$. There
are two equivalent approaches in the literature. The first one is
(a) to stay in the coordinate space that we begin, with ``old"
coordinates $(x,y) \in [0,1] \times [0,1]$, but study the solutions
of a deformed Hamiltonian. This Hamiltonian is the usual local
Hamiltonian in ``new" coordinates $(x',y')$ but now expressed in
terms of the old coordinates using the coordinate transformation $
x' + i y' = (\frac{V}{\tau_2})^{1/2} (x + \tau y)$, associated with
the deformation. The usual (periodic) boundary conditions are
applied. The second approach is (b) to apply the deformed boundary
conditions i.e. a new condition\cite{HR} in the direction of $\tau$
and stay with the same local (undeformed) Hamiltonian. Both
approaches lead to the same deformed ground state $\Psi$ which
should be used in the formula for Hall viscosity with the Berry
curvature:
\begin{equation}
\eta^{A} = 2 \;Im\langle \partial_{\tau_1} \Psi | \partial_{\tau_2}
\Psi \rangle,
\end{equation}
calculated at the reference point $V = 1$ and $\tau = i$. To recover
the physical units we should multiply with $\frac{\hbar}{L_{x}
L_{y}}$ where $L_{x}$ and $L_{y}$ are in a general case the lengths
associated with a rectangular system. In the following we will study
such a general geometry in which the deformations from a rectangle
with lengths $L_{x}$ and $L_{y}$ are made - see Figure 2.

\begin{figure}[htb]
\centering
\includegraphics[scale=0.4, angle=0]{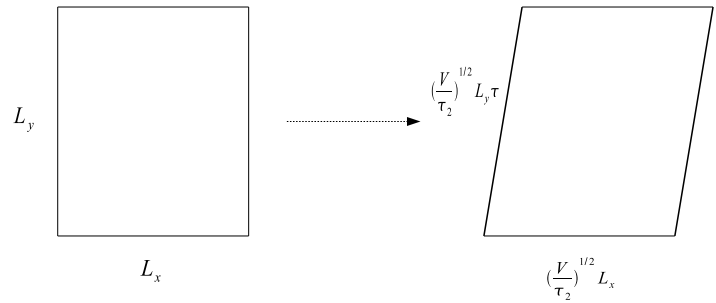}
\caption{Deformed rectangle} \label{fig2}
\end{figure}

 If we consider a gapped system of non-interacting electrons that
fill the LLL to get the Hall viscosity we have to sum the
contributions from each single particle state in the LLL. The wave
functions that describe the way how a single particle wave function
is changed as the geometry of the finite-volume system is varied
(Figure 2) are
\begin{equation}
\Psi_{j} = \sum_{k = - \infty}^{+ \infty} \exp\{ i \frac{(X_j + k
L_y) x}{l_{B}^2} + i \;\tau \;\frac{(X_j + k L_y - y)^2}{2
l_{B}^2}\}, \;\;\;\; X_j = \frac{2 \pi l_B^2 j}{L_x}. \label{twf}
\end{equation}
We did not include the normalization of each wave function that is
labeled  by an integer $ j = 0, \ldots, N_s - 1$, where $N_s =
\frac{L_x L_y}{2 \pi l_B^2}$ i.e. the number of flux quanta through
the system; in our case this number is equal to the number of
electrons. The coordinates $x$ and $y$ are old coordinates and the
wave functions satisfy the ordinary periodic BC (PBC) in
$x$-direction:
\begin{equation}
\Psi_j(x + L_x) = \Psi_j(x),
\end{equation}
and the magnetic BC in $y$- direction:
\begin{equation}
\Psi_j(y + L_y) = exp\{i \frac{L_y x}{l_B^2}\} \Psi_j(y).
\end{equation}
We will relax\cite{TVjcm} the demand for the magnetic BC
 (or we may take large $L_y$ limit \cite{comment_limit} in Eq.(\ref{twf})).
The wave function becomes simpler in this cylinder geometry:
\begin{equation}
\Psi_j = \exp\{i \frac{X_j x}{l_B^2} + i \tau \frac{(X_j - y)^2}{2
l_B^2}\} \times \frac{(\tau_2)^{1/4}}{(l_B \sqrt{\pi})^{1/2}},
\label{cyf}
\end{equation}
where we included the normalization. We have a set of orthonormal
wave functions which can be used for the calculation of the Berry
curvature as a sum of contributions of each single particle state.
As we prove \cite{lev} in the Appendix A,  if the wave function is
non-analytic i.e. non-holomorphic in $\tau$ variable only in its
normalization i.e. can be expressed as $\Psi = \frac{1}{\sqrt{Z}}
f(\tau,x,y)$, its contribution to the Hall viscosity is
\begin{equation}
\frac{\hbar}{L_x L_y} \times \frac{1}{2} (\frac{\partial^2}{\partial
\tau_1^2} + \frac{\partial^2}{\partial \tau_2^2}) \ln Z.
\label{form2}
\end{equation}
Specifying to our set, Eq.(\ref{cyf}), the sum of all contributions
is
\begin{equation}
\eta^A = \frac{\hbar n}{4}. \label{value}
\end{equation}
Therefore we recovered the well-known result (for $\nu = 1$ QHE)
using the cylinder geometry and it will be the same even if we were
applying the so-called ``thin-torus limit" i.e. cylinder limit for
which $L_x \rightarrow 0$.

\section{Hall viscosity of free Fermi gas}

Classically and in the adiabatic response theory it is expected
\cite{Avron}  that the system with time-reversal symmetry does not
have the asymmetric (Hall) viscosity. We study the asymmetric
viscosity of the free Fermi gas in the following. We will use the
Berry curvature formula to calculate the Hall viscosity even for
this system assuming that in the adiabatic response, when we probe a
finite fraction i.e. small finite system, the ground state stays
non-degenerate as $\tau$ is varied at least for a small interval
before a reconfiguration of the Fermi surface. In the case of free
Fermi gas the tiny gap $\Delta \sim \frac{1}{L^2}$, where $L$ is the
length of the system, $L = max\{L_x, L_y\}$ keeps the filling of the
Fermi see intact at least for values of $V, \tau_1,$ and $\tau_2$ in
the neighborhood of $V =1$ and $\tau = i$. Therefore because the
Hall viscosity is the Berry curvature at a specific point in the
parameter space and not an integral of it in the same space the
demand for the ground state being non-degenerate can be relaxed to
the same requirement in the neighborhood of the unperturbed point.
In fact the linear response theory leads to the Berry curvature
formula \cite{TVjcm}.

In our case of the free Fermi gas we need a small enough system. We
will adopt the approach in Ref. \onlinecite{Avron} and study the
deformed Hamiltonian:
\begin{equation}
H = -\frac{1}{V \tau_2} [|\tau|^2 \partial_x^2 - 2 \partial_x
\partial_y + \partial_y^2], \label{defham}
\end{equation}
on space $(x,y) \in [0, L_x] \times [0, L_y]$, with periodic
boundary conditions. We seek the solutions in the form:
\begin{equation}
\exp\{i k_x x' + i k_y y'\} = \exp\{i(k_x x + k_x \tau_1 y + k_y
\tau_2 y) \sqrt{\frac{V}{\tau_2}}\},
\end{equation}
as we demand that locally we have the same equation irrespective
whether we work with ``old" or ``new" coordinates. The eigenvalues
are $\epsilon(\mathbf{k}) \sim k_x^2 + k_y^2$. But the demand for
PBCs and orthogonality leads to
\begin{equation}
k_x = \frac{2 \pi}{L_x} \sqrt{\frac{\tau_2}{V}} m \;\;\; {\rm and}
\;\;\; k_y = 2 \pi \frac{1}{\sqrt{V \tau_2}} ( \frac{n}{L_y} -
\frac{m}{L_x} \tau_1),
\end{equation}
where $m$ and $n$ are integers. Therefore the quantized energy
levels are:
\begin{equation}
\epsilon(\mathbf{k})= \epsilon(n, m) = \frac{(2 \pi)^2}{V \tau_2}
[\tau_2^2 \frac{m^2}{L_x^2} + ( \frac{n}{L_y} - \frac{m}{L_x}
\tau_1)^2]. \label{disp}
\end{equation}
Though the deformation $\tau$ modifies the eigenvalues, eigenstates
are independent of it, and this leads to zero value for Berry
curvature and the Hall viscosity as expected in a time-reversal
invariant system. As we recovered the result that is valid for a
system of any size, we will use the obtained description and
formulas even in the large $N$ limit in the following.

\section{Hall viscosity of Fermi-liquid-like state}

We will consider the bosonic Fermi-liquid-like state at the filling
factor $\nu = 1$ because we expect that the conclusions will not
depend on the kind of the Laughlin-Jastrow factor in the wave
function. Therefore we study wave function:
\begin{equation}
\Psi = \Psi_L^{\nu = 1} \times Det(\exp\{i \vec{k}_i \vec{r}_j\}),
\label{flls}
\end{equation}
which is not normalized, and $\Psi_L^{\nu = 1}$ is the Vandermonde
determinant and $Det(\exp\{i \vec{k}_i \vec{r}_j\})$ the Slater
determinant of free waves. Therefore we study the unprojected to LLL
wave function. We will assume the following evolution of the wave
function under deformation $\tau$: Each factor will evolve according
the deformed single-particle Hamiltonians: (a) the one with magnetic
field as in Avron et al. with magnetic boundary conditions in the
case of the evolution of the part that ``sees" the magnetic field
i.e. Vandermonde determinant and (b) the Hamiltonian given in
Eq.(\ref{defham}) with PBCs that governs the evolution of the part
with plane waves. Therefore we assume separate evolutions that we
know very well. As we study the small deformations of a rectangular
system and plane waves do not depend on it, the most important
question is what is the shape of the Fermi sea of the unperturbed
finite Fermi system in a rectangular geometry. This is a difficult
question though we believe that in the thermodynamic limit the Fermi
sea will assume its circular, isotropic shape. But if we stay with
the shape of a rectangle even in the thermodynamic limit, as we
should do just as in the Laughlin case when we apply the Berry
curvature formula \cite{TVjcm} the question is still there
\cite{berg}. It has to be resolved only by studying the full
interacting system in the LLL. Here we will be studying (A) a
limiting case of thin torus (cylinder), (B) a system with
rectangular shape of its Fermi surface and then reach conclusions in
the case (C) which is an isotropic, circular Fermi surface.

Before that we will analyze the Berry curvature formula for the wave
function in Eq.(\ref{flls}) with the assumed evolution in general
terms. Because the part with Slater determinant of free waves does
not depend on $\tau$, the expression for the Hall viscosity is again
\begin{equation}
\eta_A = \frac{\hbar}{L_x L_y}\; \frac{1}{2}\; \Delta ln Z,
\end{equation}
where $Z$ is the norm of the wave function in Eq.(\ref{flls}) and
the derivatives are calculated at $V = 1$ and $\tau = i$.

\subsection{Thin cylinder limit}
The system can be viewed as in Fig. 3
\begin{figure}[htb]
\centering
\includegraphics[scale=0.3, angle=0]{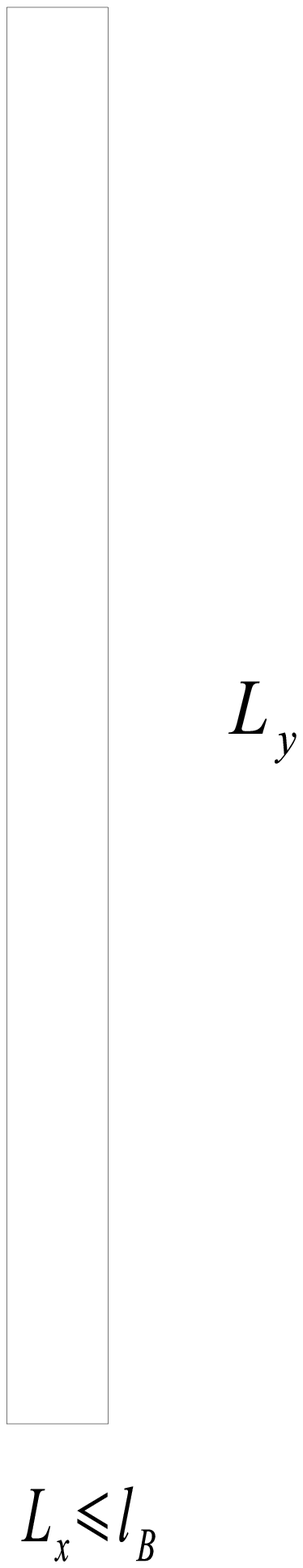}
\caption{Thin rectangle limit} \label{fig3}
\end{figure}

Then because of the PBC in $x$-direction we have the quantization of
the momentum as before
\begin{equation}
X_j = \frac{2 \pi l_B^2}{L_x} j,
\end{equation}
$j = 0,\ldots, N_s - 1$ where $N_s = \frac{L_x L_y}{2 \pi l_B^2}$.
Now we take $L_x \rightarrow 0$ limit along $L_y \rightarrow \infty$
to keep $N_s$ constant. For the Fermi-liquid-like state that means
that the neutral fermions in the $k$-space will form a line along
$y$ direction with two Fermi-points instead of a circle (line) for a
Fermi surface. In real space that is described by the following wave
function,
\begin{equation}
\prod_{i<j} \sin\{\frac{\pi}{L_y} (y_i - y_j)\}, \label{onedfl}
\end{equation}
where we assumed an odd number of electrons. Notice that there is no
$x$-dependence. Therefore when we ask for the norm of the complete
wave function (with the Laughlin-Jastrow factor at $\nu = 1$
-Vandermonde determinant) we get
\begin{equation}
Z = \prod_{i=1}^N \int dy_i \sum_{\sigma \in S_N} \exp\{- \tau_2
(y_i - k_{\sigma(i)})^2\} \prod_{k<l} \sin^2\{\frac{\pi}{L_y}(y_k -
y_l)\}.
\end{equation}
Under translations of $y$-variables:
\begin{equation}
Z = \prod_{i=1}^N \int dy_i  \exp\{- \tau_2 y_i^2\} \sum_{\sigma \in
S_N} \prod_{k<l} \sin^2\{\frac{\pi}{L_y}(y_k - y_l + k_{\sigma(k)} -
k_{\sigma(l)})\}.
\end{equation}
Due to the Gaussian factors in $y$-integration for $\tau_2 \sim 1$
we can assume that relevant values of $y$'s in the product are $y_i
\lesssim l_B, \forall i \in [1, N]$. Because $|k_{\sigma(k)} -
k_{\sigma(l)}| \geq \frac{2 \pi l_B^2}{L_x}$, when $L_x \rightarrow
0$ we can neglect the presence of $y$'s in the sine functions and
due to the scalings, $ \tau_2 y_i \rightarrow y_i$, we recover the
result for the Hall viscosity identical to the integer quantum Hall
effect at $\nu = 1$.

\subsection{The system with a rectangular shape of its Fermi
surface}

The Fermi gas with a rectangular shape of its Fermi surface may be
rather artificial but as we already discussed (a) this shape may
appear in small systems with rectangular boundaries and (b) the
conclusions reached and constructions applied to this system will
serve as a stage for discussing the problem with rectangular
boundaries in the thermodynamic limit and circular shape of the
Fermi surface.

Let us assume that we have a Fermi surface of a rectangular shape
where, for simplicity, we take that the length and width are the
same and proportional to $\sqrt{N} \in \mathbb{Z}  $. (To retain
PBCs (instead of antiperiodic BCs) we may demand that $\sqrt{N}$ is
an odd number.) The ground state function of the ideal gas has to be
an eigenvector under inversion symmetry: $ y_i \rightarrow - y_i $
and $ x_i \rightarrow x_i,  \forall i$ ( or $ y_i \rightarrow  y_i $
and $ x_i \rightarrow - x_i, \forall i$) and that constrains its
form to two possibilities: (1)
\begin{equation}
{\cal A} \Bigg\{ \prod_{\rm over\; slices\; in\; k\; space}
\Bigg[\prod_{i<j; i,j \in {\rm slice}} \sin\{\frac{\pi}{L_y}(y_i -
y_j)\} \cos\{\frac{\pi}{L_x}(x_i -
x_j)\}\Bigg]\Bigg\},\label{constr}
\end{equation}
where slices are lines in $k$ - space, along $k_x$ and $k_y$
direction, of length $\sqrt{N}$ each and to each one is assigned
$\sqrt{N}$ number of particles, see Figure 4.
\begin{figure}[htb]
\centering
\includegraphics[scale=0.3, angle=0]{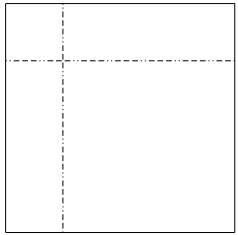}
\caption{Rectangular Fermi surface and two slices with the same
group of particles} \label{fig4}
\end{figure}
 For a fixed $\sqrt{N}$
number of particles we have two slices or lines symmetrically
positioned in $k$ space around $ k_x = - k_y$ line (Figure 4). So as
a first step we divide particles in $\sqrt{N}$ slices (lines,
groups), and at the end we antisymmetrize $({\cal A})$ that
construction in the curly brackets, which represents a particular
division into $\sqrt{N}$ groups. See an example with 4 particles in
Appendix B. (We introduced slicing in $k$ space although at this
point it seems redundant - only division in $\sqrt{N}$ groups and
later antisymmetrization  is all that is in Eq.(\ref{constr});
slicing in $k$ - space is helpful to introduce and analyze  more
general Fermi surfaces as we will see later on.)

The second possibility (2) is with $x$'s and $y$'s interchanged. If
the width and length are not the same , for example $L_y > L_x$ then
for a single slice of $\sqrt{N} \frac{L_y}{L_x}$ integers (integers
denote particles of the particular slice or group), $S_y$, we have
to symmetrize, in addition, smaller slice of $ S_x \subset S_y$
integers, with $\sqrt{N} \frac{L_x}{L_y}$ of them, i.e.
\begin{equation}
\prod_{i<j; i,j \in S_y} \sin\{\frac{\pi}{L_y}(y_i - y_j)\} S\Big
\{\prod_{k<l; k,l \in S_x}\cos\{\frac{\pi}{L_x}(x_k - x_l)\} \Big
\},
\end{equation}
so that $x$ - part is also symmetric under permutations inside
$S_y$.

Then our norm, i.e. $Z$, for the compressible quantum Hall state at
$\nu = 1$ becomes a sum of terms, each representing two fixed
permutations $\sigma, \sigma'$ of $N$ integers as in the following,
\begin{eqnarray}
 \prod_{i=1}^N \int dy_i  \prod_{l=1}^N \exp\{- \frac{\tau_2}{2} (y_l
- k_{\sigma(l)})^2\}&& \prod_{p=1}^N \exp\{- \frac{\tau_2}{2} (y_p -
k_{\sigma'(p)})^2\}\times \nonumber \\\prod_{\rm over \; slices}
&&\Big\{\prod_{k<l;k,l \in {\rm slice}} \sin \{\frac{\pi}{L_y}(y_k -
y_l)\}\Big\} \prod_{\rm over \; slices}' \Big\{\prod_{p<q;p,q \in
{\rm slice}} \sin \{\frac{\pi}{L_y}(y_p - y_q)\}\Big\}, \label{term}
\end{eqnarray}
where we suppressed (did not write) the part that corresponds to
$x$- integration. (Note now $\sigma \neq \sigma'$ in general due to
tha $x$- dependence of the wave function describing Fermi sea.) Now
we can shift each $y_i$, by $ (k_{\sigma(i)} + k_{\sigma'(i)})/2
\equiv \bar{k}_i$ and estimate how the scaling with $\tau_2$ can be
affected with the Fermi sea part. The factors that come out,
$\exp\{- \frac{\tau_2}{4} (k_{\sigma(i)} - k_{\sigma'(i)})^2\}$,
(besides the Gaussians in $y$'s) will suppress the contributions of
terms, Eq.(\ref{term}), for which $\sigma$ and $\sigma'$ differ too
much and in the following we will assume $|\sigma(i) - \sigma'(i)|
\ll \frac{L_x}{l_B}$. With this in mind we concentrate on a single
slice:
\begin{equation}
\prod_{k<l;k,l \in {\rm slice}} \sin\{\frac{\pi}{L_y}(y_k - y_l +
\bar{k}_{k} - \bar{k}_{l})\}. \label{slice}
\end{equation}
Now we ask again the question when $|\bar{k}_{k} - \bar{k}_{l}|
\lesssim l_B$. Because in this case $L_x$ is not small we can have $
\bar{k}_{i} \lesssim l_{B}$ for $ \sigma(i) \lesssim
\frac{L_x}{l_B}$ and an estimate can be that this can happen for all
pairs $\sigma(l)$ and $\sigma(k)$ in Eq.(\ref{slice}) for which
$\sigma(l),\sigma(k) \lesssim \frac{L_x}{l_B}$ and we might think
that there are $N_p \approx \frac{L_x}{l_B} (\frac{L_x}{l_B} - 1)/2$
of them. But this is an overestimate for the construction in
Eq.(\ref{constr}) because by making division in slices we do not
have the factor $\sin \{\frac{\pi}{L_y}(y_k - y_l)\}$ for each pair
of particles. As we do not have as many pairs as $N (N - 1)/2$, but
because of slicing only around $\sim L_x \times L_y^2/ l_B^3({\rm
or}\;\; N^{3/2} \frac{L_y}{L_x})$, the $N_p$ should be reduced
\cite{comment_prob} by $L_x$ and therefore is not of the order of
$N$, which would pose a problem in the scaling argument for the Hall
viscosity and influence its final value. Therefore we argued that we
can model the contribution of each term as in Eq.(\ref{term}) with
$\sigma \approx \sigma'$ as
\begin{equation}
\sim \frac{1}{\tau_2^{N/2 + \alpha}} I(L_x, L_y),\label{form}
\end{equation}
where $ \alpha \lesssim \frac{L_x}{l_B}$. Even if the specific value
of $\alpha$ depends on the choice of grouping of particles for
slicing in Eq.(\ref{constr}), by extracting a leading contribution
in $N$ we can recover the same result for the Hall viscosity as
before. We in fact are taking the large $N$ limit before the limit
$\tau \rightarrow i$, which is allowed \cite{comment} but implies
the same value of the Hall viscosity only in this limit. More
precisely, if we do not assume the effective reduction of each
$\sin\{\frac{\pi}{L_y}(y_k - y_l + \bar{k}_{k} - \bar{k}_{l})\}$ to
either $\sin\{\frac{\pi}{L_y}(y_k - y_l)\} \sim \frac{\pi}{L_y}(y_k
- y_l)$ or $\sin\{\frac{\pi}{L_y}(\bar{k}_{k} - \bar{k}_{l})\}$, in
the end the term in Eq.(\ref{term}) can be expressed as a series
with each member of the form as in Eq.(\ref{form}), where again $
\alpha \lesssim \frac{L_x}{l_B}$, and the argument follows. This is
possible because for any $|\bar{k}_{k} - \bar{k}_{l}| \lesssim l_B$
and, as we have due to the shifts and Gaussians $|y_{k} - y_{l}|
\lesssim l_B$, we can approximate
\begin{equation}
\sin\{\frac{\pi}{L_y}(y_k - y_l + \bar{k}_{k} - \bar{k}_{l})\}
\approx \frac{\pi}{L_y}(y_k - y_l + \bar{k}_{k} - \bar{k}_{l}),
\end{equation}
and an expansion in the differences in $y$'s i.e.
$\frac{\pi}{L_y}(y_k - y_l$) follows.

\subsection{The system with circular Fermi surface}

Our expectation is that the composite fermions will make an
isotropic, circular Fermi surface even in the thermodynamic limit of
the system with rectangular boundaries. Nevertheless in this case we
have to demand that the ground state of the system retains the
inversion symmetry of the system in its neutral sector, i.e. that
the Fermi part of the ground state wave function is an eigenvector
under $y_i \rightarrow - y_i$ and $ x_i \rightarrow x_i$, $\forall
i$, transformation. The rectangular shape is a feature of the system
on which the shear transformation is applied. Such a ground state
wave function can be constructed by a generalization of the
construction in the previous case (B) given in Eq.(\ref{constr}).
Now the two slices along $k_x$ and $k_y$ direction are as in Figure
5.
\begin{figure}[htb]
\centering
\includegraphics[scale=0.4, angle=0]{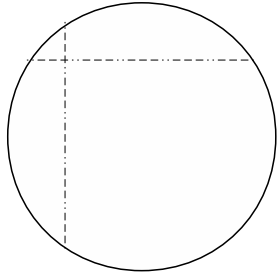}
\caption{Circular Fermi surface and two slices with the same group
of particles} \label{fig5}
\end{figure}
A group of particles, its number equal to the length of the two
slices is assigned to them. As we sweep the whole circle with these
two slices we make a certain division of all particles into groups
that correspond to slices and therefore, to make the wave function
antisymmetric under particle exchange, we need an overall
antisymmetrization as in Eq.(\ref{constr}). The same arguments as in
the case with rectangular Fermi surface can be applied here and lead
to the conclusion that the Hall viscosity of the Fermi-liquid-like
quantum Hall state is unaffected by the presence of the Fermi see in
the ground state function. Namely all estimates that we did in the
rectangular case will be modified by geometrical factors that will
not affect the conclusion on the leading behavior in the
thermodynamic limit. To illustrate what we mean by geometrical
factors let us consider a Fermi surface that is a square with
$\sqrt{N}$ length of each side. In that case the number of pairs is
$ \sqrt{N} \times \sqrt{N}(\sqrt{N} - 1)/2 $, but in the case of a
circle, which delineates the same volume equal to $N$, we will have
for the same quantity $ \frac{16}{\pi^{3/2}} N^{3/2} + 4 N$ i.e. the
same leading behavior $\sim N^{3/2}$ up to a numerical - geometrical
factor.

\section{The Hall viscosity at $\tau \neq i$}

The most important question when considering the question of the
Hall viscosity for the Fermi-liquid-like state is whether it is
dependent on $\tau$ or maybe it is independent of the geometry
$(\tau)$, which is a remarkable property of the integer quantum Hall
state\cite{Avron}  at $\nu = 1$ and other quantum Hall states
\cite{ReadV} that exhibit Hall conductance plateaus.

When considering arbitrary $\tau$ we have to start with the deformed
Fermi surface as it follows from the deformed dispersion relation in
Eq.(\ref{disp}). To simplify the notation we will take $L_x = L_y =
L$ or, in general, that $m$ and $n$ carry factors connected with the
lengths and may not be integers. Therefore we write the dispersion
relation $\epsilon(\tau)$ as
\begin{equation}
\epsilon(\tau) = \frac{(2 \pi)^2}{V L^2 \tau_2} [ \tau^2_2 m^2 + (n
- \tau_1 m)^2],
\end{equation}
or if we absorb the scaling factor $ f(\tau) = \frac{(2 \pi)^2}{V
L^2 \tau_2}$ and define $e(\tau)$ as $\epsilon(\tau) = f(\tau)
e(\tau)$ we may focus on the dispersion relation expressed as
\begin{equation}
e(\tau) =  \tau^2_2 m^2 + (n - \tau_1 m)^2.
\end{equation}
The equation $\epsilon_F = e_F f$ where $\epsilon_F$ is the fermi
energy defines the (deformed and scaled) Fermi surface i.e.
\begin{equation}
e_F \equiv e =  \tau^2_2 m^2 + (n - \tau_1 m)^2,
\end{equation}
which is illustrated in Figure 6.
\begin{figure}[htb]
\centering
\includegraphics[scale=0.45, angle=0]{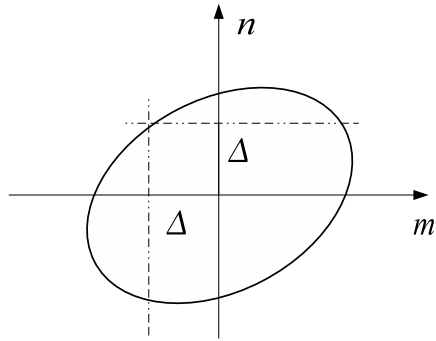}
\caption{Deformed Fermi surface and two slices with the same group
of particles} \label{fig6}
\end{figure}

We find that the maximum values of $m$ and $n$ (that belong to
points on the Fermi surface) are $ m_{max} =
\frac{\sqrt{e}}{\tau_2}$ and $ n_{max} = \frac{\sqrt{e}}{\tau_2}
|\tau|$, respectively. And for $m = - \Delta$ we have corresponding
$n = - \tau_1 \Delta \pm \sqrt{e - \tau_2^2 \Delta^2}$ and for $ n =
\Delta$ we have $m =(\tau_1 \Delta \pm \sqrt{e - \tau_2^2
\Delta^2})/(|\tau|^2)$. This implies that if we keep $|\tau| = 1$ we
would have the same length for corresponding two slices along $k_x$
and $k_y$ directions that we introduced before. To simplify the
discussion in the following we will assume that this is the case
i.e. that due to $|\tau| = 1$ we have the symmetry under inversion
around the axis defined by $n = \tau_1 m$.

As our deformed Hamiltonian in Eq.(\ref{defham}) has the symmetry
under simultaneous transformations $\tau_1 \rightarrow - \tau_1$ and
$y \rightarrow - y, x \rightarrow x$ or $\tau_1 \rightarrow -
\tau_1$ and $y \rightarrow  y, x \rightarrow - x$ our dispersion
relation, Eq.(\ref{disp}), has the same symmetry and the
corresponding Fermi surface as well. This symmetry has to exist in
the ground state, which has to accommodate to the deformed
rectangular shape for $\tau \neq i$. For $ \tau = i$ the symmetry
can be identified as the inversion symmetry around $x$ or $y$ axis
that has to be generalized to the case with $\tau \neq i$ for which
we need to include also $\tau_1 \rightarrow - \tau_1$
transformation. With this in mind we can come up with a ground state
wave function that will have this symmetry in the Fermi part under
simultaneous transformations in coordinate and momentum space. Using
the slice decomposition that is illustrated in Figure 6 for the
deformed Fermi surface and with the simplifying assumption $|\tau| =
1$ the Fermi part will look like
\begin{equation}
{\cal A} \Bigg\{ \prod_{\rm over\; slices\; in\; k\; space}
\Bigg[\prod_{i<j; i,j \in {\rm slice}} \exp\{i k_y^c \sum_{i \in
{\rm slice}} y_i\}\sin\{\frac{\pi}{L_y}(y_i - y_j)\} \exp\{i k_x^c
\sum_{i \in {\rm slice}} x_i\} \cos\{\frac{\pi}{L_x}(x_i -
x_j)\}\Bigg]\Bigg\},\label{dconstr}
\end{equation}
where slices in $k_x$ and $k_y$ direction correspond in the manner
of Figure 6; to each slice in $k_y$ corresponds the slice of the
same length in $k_x$ corresponding to the same group of particles.
The exponentials with $k_x^c$ and $k_y^c$ carry the momentum
$\vec{k}^c = (k_x^c, k_y^c)$, which is due to the deformation of the
Fermi surface and the absence of the inversion symmetries around
$k_x$ and $k_y$ axis. The momentum $\vec{k}^c$ lies along the new
symmetry line i.e. $k^c_y = \tau_1 k_x^c$ and represents the
momentum of the center of the mass of the particles that belong to
the particular slice. ${\cal A}$ in Eq.(\ref{dconstr}) is again the
overall antisymmetrization that will bring all possible assignments
of particles into slices in the final form of the wave function. The
construction when $x$'s and $y$'s ($k_x^c$'s and $k_y^c$'s) are
interchanged is also possible and we will discuss that case later.

The complete deformed wave function for the Fermi-liquid-like state
has the Gaussian factors of the form:
\begin{equation}
\exp\{- \frac{\tau_2}{2}(y_i - k_{\sigma(i)})^2\}
\end{equation}
that enter the integral for $Z$. If $\tau_2$ is small the $y_i$ can
fluctuate being less localized with the Gaussian. So the relevant
interval of $y_i$ values in the integral becomes larger and the
sequence of approximations for two particles, beginning with the
corresponding term in the product in the integral
(Eq.(\ref{slice})), as
\begin{equation}
 \sin\{\frac{\pi}{L_y}(y_k - y_l + \bar{k}_{k} - \bar{k}_{l})\}
 \approx \sin\{\frac{\pi}{L_y}(y_k - y_l)\} \approx \frac{\pi}{L_y}(y_k -
 y_l),
\label{oneslice}
\end{equation}
are more likely to be allowed. Each term like this will contribute $
1/\sqrt{\tau_2}$ when the scaling $ \sqrt{\tau_2} y_i \rightarrow
y_i, \forall i$ is applied. If, for small enough $\tau_2$, we assume
that for each pair we can do this approximation, in addition to the
overall exponent, which we get by the change of variables in the
$y$-integration, of order $N$, we will get another contribution of
order $N^{3/2}$ that would lead to the divergence of the Berry
curvature and therefore for finite $\tau_2$ of the Hall viscosity.
This is certainly an overestimate but the possibility of divergence
seems lurking. Applying arguments similar to the one in Section IV B
we come to an estimate that the number of relevant pairs is around
$\sim \frac{L_x}{l_B} \frac{1}{\tau_2}$. Therefore only for strong
deformations for which $\tau_2 \sim \frac{1}{\sqrt{N}}$, we may
expect the departure of the value for the Hall viscosity from the
one of Laughlin states. These arguments can not be precisely
quantified but suggest that the Hall viscosity of the
Fermi-liquid-like state may depend on the value of $\tau$ for very
large deformations. But as we do apply the large $N$ limit to
recover the Laughlin state value for the Hall viscosity as $\tau
\rightarrow i$ and because here relevant $\tau_2$ is of the order of
$\frac{1}{\sqrt{N}}$, we can expect the same Hall viscosity value in
the same limit for any finite $\tau_2$. Therefore the feature of the
Laughlin states that their Hall viscosity is independent of $\tau$
may stand as a reflection of their true topological nature due to
the comparison with the Fermi-liquid-like state that can recover the
same value only in large $N$ limit.

The precise estimate how the Hall viscosity depends on $\tau$ (in
the case of the Fermi-liquid-like state) is hard to get also because
of the exponentials with $\vec{k}^c$ that carry dependence on $y_i$.
(We have to keep in mind that the scaling $ \sqrt{\tau_2} y_i
\rightarrow y_i, \forall i$ is a purely mathematical transformation
of variables under the $Z$ integral and does not affect
$k$-variables.) In the argument above we assumed $\sum_{i \in S} y_i
\approx 0$ for each slice $S$, which might not be the case.

\section{The inversion symmetry and Hall viscosity}

For $\tau = i$ we view the inversion symmetry as the symmetry under
transformations $y_i \rightarrow - y_i$ and $ x_i \rightarrow x_i$,
$\forall i$ around $x$-axis or $y_i \rightarrow  y_i$ and $ x_i
\rightarrow - x_i$, $\forall i$ around $y$-axis. For $\tau \neq i$,
as we already noted, it can be generalized by adding $\tau_1
\rightarrow - \tau_1$ transformation. The symmetry has to be
incorporated in the ground state wave function, more precisely in
its neutral part, when we discuss the system with rectangular shape
(or deformed rectangular shape, $\tau \neq i$) and our aim is the
calculation of the Hall viscosity.

In the case of the Fermi-liquid-like state two constructions stand
out at $\tau = i$ (and their generalizations for $\tau \neq i$) for
the Fermi part
\begin{equation}
{\cal A} \Bigg\{ \prod_{\rm over\; slices\; in\; k\; space}
\Bigg[\prod_{i<j; i,j \in {\rm slice}} \sin\{\frac{\pi}{L_y}(y_i -
y_j)\} \cos\{\frac{\pi}{L_x}(x_i - x_j)\}\Bigg]\Bigg\},\;\;\;\;\;\;
(a)\label{aconstr}
\end{equation}
with the notation that we explained previously, and
\begin{equation}
{\cal A} \Bigg\{ \prod_{\rm over\; slices\; in\; k\; space}
\Bigg[\prod_{i<j; i,j \in {\rm slice}} \cos\{\frac{\pi}{L_y}(y_i -
y_j)\} \sin\{\frac{\pi}{L_x}(x_i - x_j)\}\Bigg]\Bigg\}.\;\;\;\;\;\;
(b)\label{bconstr}
\end{equation}
They are explicitly invariant under the inversion symmetry
transformations. The constructions are valid for both circular and
rectangular Fermi surface. In Appendix B we display the functions
$(a)$ and $(b)$ in the case of 4 particles. In that case it can be
easily shown that the state - construction that is $ {\cal A}\{
\sin\{\frac{\pi}{L_y}(y_1 - y_2)\} \cos\{\frac{\pi}{L_y}(y_3 -
y_4)\} \ldots\}$, is identical to zero. As the square of the
inversion symmetry is equal to identity in general we expect that
wave functions $(a)$ and $(b)$ represent two degenerate ground
states and two independent sectors of the Fermi liquid.

Throughout the paper we discussed the case $(a)$ for the
Fermi-liquid-like state and concluded that, in the thermodynamic
limit, around $\tau = i$ the Hall viscosity is equal to the one of
Laughlin states and that our expectation is that for general $\tau
\neq i$ this will still be true in the same limit. If we try similar
arguments in the case $(b)$ we can come to the expectation that, due
to the cosine functions in the dependence on $y$'s, no change in the
overall scaling with $\tau_2$ will occur and for this construction
 the Hall viscosity is independent of $\tau$ and equal
to the one of the Laughlin states.

The single particle Hamiltonian that describes the evolution of the
part of the Fermi-liquid-like state that sees magnetic field is not
invariant under the inversion symmetry and the ``true" energetics of
the problem at $\nu = 1$ (fermionic at $\nu = 1/2$) will certainly
differentiate between the two possibilities for the ground state:
with Fermi parts $(a)$ and $(b)$. We expect that the construction
with Fermi part $(a)$, irrespective whether the ground state is
non-degenerate or degenerate will make a ground state as it can
smoothly evolve from the thin torus limit and its Fermi part,
Eq.(\ref{onedfl}), when the gauge is fixed so that the Gaussians are
along $y$-axis. The construction with $(b)$ Fermi part may appear as
an additional sector.

\section{Discussion and Conclusions}

In this paper we discussed the Berry curvature calculations of the
Hall viscosity for the unprojected to the LLL wave function of the
Fermi-liquid-like state. We concluded that in the linear response,
with small deformation of the system and in the thermodynamic limit,
the Hall viscosity takes the value characteristic for the Laughlin
states (Eq.(\ref{value})). We presented arguments that the value is
the same even for general deformations in the same limit.

The preprint in Ref. \onlinecite{rrhv} appeared very recently when
we were in the process of finishing of the present paper. There the
claim is made, on the basis of the Berry curvature formula applied
to the wave functions in the LLL (or projected to a definite LL),
that at $\nu = 1/2$, irrespective whether the state is
incompressible or not, if the Hamiltonian is particle-hole
symmetric, the Hall viscosity acquires the Laughlin value. The value
is the same irrespective of the deformation ($\tau$).  Though our
analysis is on the unprojected (to the LLL) wave function of
Fermi-liquid-like state, we agree on the value of the Hall viscosity
for the state at $\nu = 1/2$ in the thermodynamic limit. For a
general $\tau$ it is surprising that the same value of the Hall
viscosity is maintained in the LLL \cite{rrhv,hald}, and somehow has
to be reconciled with the expected quantization in the
incompressible states. A way out is to claim that the
Fermi-liquid-like state has the dissipative (symmetric) viscosity
nonzero \cite{ReadV}, but still the quantization of the Hall
viscosity for the compressible state in the LLL undermines our
expectation that in the Hall viscosity we have yet another
characteristic of the incompressible quantum Hall states that is
quantized i.e. has a constant value under small changes
(perturbations) of the Hamiltonian. (In other words even gapless
phases may have an invariant such as the Hall viscosity.)

On the other hand the Fermi-liquid-like state and the firmly
established phase \cite{hlr} at $\nu = 1/2$ may be viewed as some
kind of a critical state where effective particle and hole physics
and two Jain's sequences of particle states (from $\nu = 1/3$) and
hole states (from $\nu = 2/3$) meet. The situation is somewhat
similar or reminiscent of the graphene and the critical behavior of
the neutrality point of the Dirac fermions \cite{sach}. Nevertheless
it looks conclusive \cite{rrhv} no critical behavior in the case of
the state at $\nu = 1/2$ and the Hall viscosity is expected.

Our study shows that the Hall viscosity of the unprojected
Fermi-liquid-like state at general $\tau$ may deviate  from the
Laughlin state value for finite number of particles. Maybe the
behavior for finite number of particles can not be explained by
non-interacting or weakly-interacting CF physics if we stay in the
LLL as it involves higher LL physics i.e.  all energy scales, which
may reflect its critical nature.

In the adiabatic transport theory that we apply our basic assumption
is that flux changing excitations are not relevant or  higher in
energy for the calculation of the Hall viscosity. The result of Ref.
\onlinecite{rrhv} seems to give credence to this approach. Within
assumptions made, we  established that the Fermi-liquid-like state
in the thermodynamic limit in the linear response has the value of
Hall viscosity equal to the value of Laughlin states. We hope that
our analysis will help further elucidation of the problem and the
search for the final answer.

\section*{Acknowledgments}
The author thanks N. Read for his comments. This work was supported
by the Serbian Ministry of Science under Grant No. 141035.

\appendix
\section{}
 We will prove the formula Eq.(\ref{form2}) for the ground state
function that is holomorphic in $\tau$ variable except for the
normalization, which is the case also with the Laughlin wave
function. The normalized wave function is $\Psi_{0} =
\frac{\Psi_L}{\sqrt{Z}}$ where $\Psi_L$ depends on particle
coordinates and $\tau$ only. We want to calculate:
\begin{equation}
Im\; \frac{\partial \langle \Psi_0|}{\partial \tau_1} \frac{\partial
| \Psi_0 \rangle}{\partial \tau_2}.
\end{equation}
First we have for $\tau_i; i = 1,2$
\begin{equation}
\frac{\partial | \Psi_0 \rangle}{\partial \tau_i} =
\frac{1}{\sqrt{Z}} \frac{\partial | \Psi_L \rangle}{\partial \tau_i}
- \frac{1}{2} \frac{\partial ln Z}{\partial \tau_i} |\Psi_0 \rangle.
\end{equation}
Therefore
\begin{equation}
Im\; \frac{\partial \langle \Psi_0|}{\partial \tau_1} \frac{\partial
| \Psi_0 \rangle}{\partial \tau_2} = Im \{ \frac{1}{Z}
\frac{\partial \langle \Psi_L|}{\partial \tau_1} \frac{\partial |
\Psi_L \rangle}{\partial \tau_2} - \frac{1}{2} \frac{1}{\sqrt{Z}}
\frac{\partial ln Z}{\partial \tau_1} \langle \Psi_0|\frac{\partial
| \Psi_L \rangle}{\partial \tau_2} - \frac{1}{2} \frac{1}{\sqrt{Z}}
\frac{\partial ln Z}{\partial \tau_2}  \frac{\partial \langle
\Psi_L|}{\partial \tau_1} |\Psi_0 \rangle\}.
\end{equation}
$|\Psi_L \rangle$ is holomorphic in $\tau$, therefore:
\begin{equation}
\frac{\partial |\Psi_L \rangle}{\partial \bar{\tau}} =
\frac{\partial |\Psi_L \rangle}{\partial \tau_1} + i \frac{\partial
|\Psi_L \rangle}{\partial \tau_2} = 0 \;\;\; {\rm and}\;\;\;
\frac{\partial |\Psi_L \rangle}{\partial \tau} = \frac{\partial
|\Psi_L \rangle}{\partial \tau_1} - i \frac{\partial |\Psi_L
\rangle}{\partial \tau_2} = 0.
\end{equation}
Then\\ (1)
\begin{equation}
\frac{\partial \langle \Psi_L|}{\partial \tau} \frac{\partial |
\Psi_L \rangle}{\partial \bar{\tau}} = \frac{\partial \langle
\Psi_L|}{\partial \tau_1} \frac{\partial | \Psi_L \rangle}{\partial
\tau_1} + \frac{\partial \langle \Psi_L|}{\partial \tau_2}
\frac{\partial | \Psi_L \rangle}{\partial \tau_2} + i \frac{\partial
\langle \Psi_L|}{\partial \tau_1} \frac{\partial | \Psi_L
\rangle}{\partial \tau_2} - i \frac{\partial \langle
\Psi_L|}{\partial \tau_2} \frac{\partial | \Psi_L \rangle}{\partial
\tau_1} = 0.
\end{equation}
It follows that
\begin{equation}
\frac{\partial \langle \Psi_L|}{\partial \tau_1} \frac{\partial |
\Psi_L \rangle}{\partial \tau_1} + \frac{\partial \langle
\Psi_L|}{\partial \tau_2} \frac{\partial | \Psi_L \rangle}{\partial
\tau_2} - 2 Im \; \frac{\partial \langle \Psi_L|}{\partial \tau_1}
\frac{\partial | \Psi_L \rangle}{\partial \tau_2} = 0, \label{one}
\end{equation}
(2)
\begin{equation}
\frac{\partial ln Z}{\partial \tau_2} = \frac{1}{Z} (\langle \Psi_L|
\frac{\partial |\Psi_L \rangle}{\partial \tau_2} + \frac{\partial
\langle \Psi_L|}{\partial \tau_2} |\Psi_L \rangle) = -
\frac{1}{\sqrt{Z}} 2\; Im \; \langle \Psi_L| \frac{\partial |\Psi_L
\rangle}{\partial \tau_1}, \label{two}
\end{equation}
(3)
\begin{equation}
\frac{\partial ln Z}{\partial \tau_1} = \frac{1}{Z} (\langle \Psi_L|
\frac{\partial |\Psi_L \rangle}{\partial \tau_1} + \frac{\partial
\langle \Psi_L|}{\partial \tau_1} |\Psi_L \rangle) =
\frac{1}{\sqrt{Z}} 2\; Im \; \langle \Psi_L| \frac{\partial |\Psi_L
\rangle}{\partial \tau_2}.\label{three}
\end{equation}
From (\ref{two}) and (\ref{three}) it follows
\begin{equation}
Im\; \frac{\partial \langle \Psi_0|}{\partial \tau_1} \frac{\partial
| \Psi_0 \rangle}{\partial \tau_2} = Im \{ \frac{1}{Z}
\frac{\partial \langle \Psi_L|}{\partial \tau_1} \frac{\partial |
\Psi_L \rangle}{\partial \tau_2}\} - \frac{1}{4} (\frac{\partial ln
Z}{\partial \tau_1})^2 - \frac{1}{4} (\frac{\partial ln Z}{\partial
\tau_2})^2.
\end{equation}
Because
\begin{equation}
\frac{\partial^2 ln Z}{\partial \tau_i^2} = \frac{\partial}{\partial
\tau_i} \frac{1}{Z} \frac{\partial Z}{\partial \tau_i} = -
\frac{1}{Z^2} ( \frac{\partial Z}{\partial \tau_i})^2 + \frac{1}{Z}
\frac{\partial^2 Z}{\partial \tau_i^2},
\end{equation}
and because of (\ref{one})
\begin{equation}
\Delta Z = \frac{\partial^2}{\partial \tau \partial \bar{\tau}} Z =
\frac{\partial \langle \Psi_L|}{\partial \bar{\tau}} \frac{\partial
| \Psi_L \rangle}{\partial \tau} = \frac{\partial \langle
\Psi_L|}{\partial \tau_1} \frac{\partial | \Psi_L \rangle}{\partial
\tau_1} + \frac{\partial \langle \Psi_L|}{\partial \tau_2}
\frac{\partial | \Psi_L \rangle}{\partial \tau_2} + 2 Im \;
\frac{\partial \langle \Psi_L|}{\partial \tau_1} \frac{\partial |
\Psi_L \rangle}{\partial \tau_2} = 4 Im \; \frac{\partial \langle
\Psi_L|}{\partial \tau_1} \frac{\partial | \Psi_L \rangle}{\partial
\tau_2},
\end{equation}
we have
\begin{equation}
Im\; \frac{\partial \langle \Psi_0|}{\partial \tau_1} \frac{\partial
| \Psi_0 \rangle}{\partial \tau_2} = \frac{1}{4} \Delta ln Z.
\end{equation}
\section{}

For four particles the construction in Eq.(\ref{constr}) or
Eq.(\ref{aconstr}) is
\begin{eqnarray}
\Psi_b &=& \sin\{\frac{\pi}{L_y}(y_1 - y_2)\}
\sin\{\frac{\pi}{L_y}(y_3 - y_4)\} \cos\{\frac{\pi}{L_x}(x_1 -
x_2)\}
\cos\{\frac{\pi}{L_x}(x_3 - x_4)\} \nonumber \\
&& -\sin\{\frac{\pi}{L_y}(y_1 - y_3)\} \sin\{\frac{\pi}{L_y}(y_2 -
y_4)\} \cos\{\frac{\pi}{L_x}(x_1 - x_3)\}
\cos\{\frac{\pi}{L_x}(x_2 - x_4)\} \nonumber \\
&& +\sin\{\frac{\pi}{L_y}(y_1 - y_4)\} \sin\{\frac{\pi}{L_y}(y_2 -
y_3)\} \cos\{\frac{\pi}{L_x}(x_1 - x_4)\} \cos\{\frac{\pi}{L_x}(x_2
- x_3)\}, \label{a}
\end{eqnarray}
and the one in Eq.(\ref{bconstr}) is
\begin{eqnarray}
\Psi_a &=& \cos\{\frac{\pi}{L_y}(y_1 - y_2)\}
\cos\{\frac{\pi}{L_y}(y_3 - y_4)\} \sin\{\frac{\pi}{L_x}(x_1 -
x_2)\}
\sin\{\frac{\pi}{L_x}(x_3 - x_4)\} \nonumber \\
&& -\cos\{\frac{\pi}{L_y}(y_1 - y_3)\} \cos\{\frac{\pi}{L_y}(y_2 -
y_4)\} \sin\{\frac{\pi}{L_x}(x_1 - x_3)\}
\sin\{\frac{\pi}{L_x}(x_2 - x_4)\} \nonumber \\
&& +\cos\{\frac{\pi}{L_y}(y_1 - y_4)\} \cos\{\frac{\pi}{L_y}(y_2 -
y_3)\} \sin\{\frac{\pi}{L_x}(x_1 - x_4)\} \sin\{\frac{\pi}{L_x}(x_2
- x_3)\}. \label{b}
\end{eqnarray}
i.e. with $x$'s and $y$'s interchanged.

The wave functions $\Psi_a$ and $\Psi_b$ can be represented by their
configurations in $k$-space. Below each configuration describes the
placements of four fermions  in the corners of a square that
correspond to allowed values of four momenta.
\begin{eqnarray}
\Psi_{a(b)}&=& +\frac{3|1}{2|4} \pm \frac{1|3}{4|2} \pm
\frac{2|4}{3|1} + \frac{4|2}{1|3} \nonumber \\
&&\mp \frac{4|1}{2|3} - \frac{1|4}{3|2} - \frac{2|3}{4|1} \mp
\frac{3|2}{1|4} \nonumber \\
&&- \frac{2|1}{3|4} \mp \frac{1|2}{4|3} \mp \frac{3|4}{2|1} -
\frac{4|3}{1|2} \nonumber \\
&&\pm \frac{4|1}{3|2} + \frac{1|4}{2|3} + \frac{3|2}{4|1} \pm
\frac{2|3}{1|4} \nonumber \\
&&+ \frac{2|1}{4|3} \pm \frac{1|2}{3|4} \pm \frac{4|3}{2|1} +
\frac{3|4}{1|2} \nonumber \\
&&\mp \frac{3|1}{4|2} - \frac{1|3}{2|4} - \frac{4|2}{3|1} \mp
\frac{2|4}{1|3}.
\end{eqnarray}


\begin{thebibliography}{99}
\bibitem{comment_limit} The limit will require additional
assumptions in our arguments and, if not stated otherwise, we take
$L_y \sim L_x$ in the following to simplfy the discussion.
\bibitem{comment_prob} The probability for two particles to pair is
$\sim l_B/L_x$.
\bibitem{comment} Although the thermodynamic limit may bring
divergences in $ \ln Z $ that grow faster than $N$ ($Z$ does not
have in general the meaning of a statistical partition function), by
decoupling the $\tau_2$ dependence in the large $N$ limit
($\frac{\alpha}{N} \sim 0$) what is left becomes independent of
$\tau_2$ and does not influence the final value for the Hall
viscosity.
\end{thebibliography}
\end{document}